 \documentstyle[11pt,paspconf,psfig]{article}

\input psfig.sty

\def\refe{\par\noindent\hangindent=1.5cm}
\keywords{gravitational lensing -- dark matter -- 
galaxies: halos}

\begin{document}

 \title{Microlensing Results and Removing the Baryonic Degeneracy}

\author{Milan M. \'Cirkovi\'c\altaffilmark{1}}
  
 \affil{Dept. of Physics \& Astronomy, SUNY at Stony Brook,
Stony Brook, NY 11794-3800, USA}

\author{Srdjan Samurovi\'c  and}

\affil{Dipartimento di Astronomia, Universit\`{a} degli Studi di Trieste,  Via Tiepolo 11, I-34131, Trieste, Italy}

\author{Vesna Milo\v sevi\'c-Zdjelar}

\affil{Dept. of Physics \& Astronomy, University of Manitoba, Winnipeg, Mb,  
R3T 2N2, Canada}



\altaffiltext{1}{Affiliated to: Astronomska Opservatorija, Volgina 7, 11000 Belgrade,
Serbia }

\begin{abstract} 
One of the biggest obstacles to our full 
understanding of the global dynamics in Milky Way and other spiral
galaxies is uncertainty with respect to the form of baryonic dark 
matter in galactic haloes. Two basic forms discussed
recently are MACHOs and various states of halo gas. We investigate constraints which could be obtained from the empirical microlensing 
optical depth on combined MACHO + gas models, and show that improved
statistics  will certainly be able to discriminate
between various such models. This has profound consequences not only 
for galactic dynamics and baryonic budget, but for investigation 
of the low-redshift Ly$\alpha$ absorption systems and general 
cosmological distribution of gas.  
\end{abstract}

\section{Constraints on the BDM, and BDM halo models}

Microlensing (ML) observations   revolutionized 
galactic halo modeling. The abundance and
properties of lenses influence the cosmological picture
of the abundance, structure and evolution of the baryonic content of
the universe.
 
Widely held  
assumptions: (I) MACHOs are baryonic, (II) Milky Way is a typical $L_\ast$ spiral galaxy,  are necessary for correct account of baryons,  comparison  with  BBNS bounds (Fields et al.~1998),   
and  comparing local mass census with total mass-energy of the universe.
 Our assumption: (III) MACHOs possess finite mass-to-light ratios, exclude dense clouds, or stellar-mass black holes from  (I).

Unknown baryonic fraction of primordial galactic haloes $f_g$ became  acute problem with the discovery of non-zero 
cosmological constant.  It can vary between 0.04 and 0.2,  depending on precise values of 
light element abundances, Hubble  constant and cosmological constant (Gates, Gyuk \& Turner~1995).

\section{Discussion}

Contribution of MACHOs to the total baryonic density, if Milky Way is typical for the present epoch,  
$\Omega_{\rm MACHO}/{\Omega_B} \simeq 0.7 \, h$ (Fields et al.~1998), 
 is based on ML surveys toward  Magellanic Clouds, still in progress. If MACHO halo extends further 
than the distance to Magellanics, $\sim 50$ kpc, the ratio also 
increases.

Inclusion of  gaseous phases into Galactic 
BDM mass budget is  based on a persistence of  Ly$\alpha$ absorption systems down to low redshifts ({\it HST\/} observations), and their proven association  with normal  galaxies  (Fukugita et al.~1998), as diffuse remnants  of huge gaseous haloes galaxies once
possessed.   With maximal absorption radius  $\sim 178\, h^{-1}$ kpc 
(Chen et al.~1998), they are   much larger than MACHO haloes,  suggesting that amount of gas
in various ionization stages around normal luminous galaxies is 
high, contrary to the conventional wisdom.

 High  baryon budget, discussed by Milo\v sevi\'c-Zdjelar, Samurovi\'c \& \'Cirkovi\'c (this
Conference) and   Jakobsen (1998), 
can not be found in the present-day IGM. We conclude that by the recent  epochs, most  baryons have been incorporated into collapsed structures  of gas of varying ionization stages,  and MACHOs.

Extending
discussion of Gates et al.~(1995) with 
gaseous phase of halo matter we establish contact between the observed  column density and physical density of gas, integrate  density over the halo volume, integrate masses of such haloes over Schechter luminosity function, and obtain three
  density profiles.   
In  model which gives the physical density {\it a priori\/}, we may  substitute the first step for integration of this density along the line-of-sight in order to compare with the observational data on the column density spatial distribution (\'Cirkovi\'c 1999).

ML results, as manifested in
optical depths and event durations, suffer from intrinsic degeneracies
which preclude construction of a realistic, 3-D model of lens 
distribution. Degeneracy could be removed by supplementing  ML data with other independent constraints from different branches of astronomy. 

The explosive advances recently made in the 
 study of gas around galaxies at low redshift promise that 
supplementing of this information to the ML data, and taking into account "high-precision era" of BBNS studies, will enable fixing the MACHO abundance and the extent of MACHO halo for $L_\ast$ galaxies. Composite BDM models may certainly aspire to be more realistic than gas-alone (Mo \& Miralda-Escud\'e 1996)   or
the MACHO-dominated (e.g.~Honma \& Kan-ya 1998).
 
\acknowledgments

Srdjan Samurovi\'c acknowledges the financial support of the University of Trieste.

\section*{References}

\refe Chen, H.-W., Lanzetta, K. M., Webb, J. K., \& Barcons, X. 1998, ApJ, 498, 77 

\refe \'Cirkovi\'c, M. M. 1999, Ap \& SS, in press

\refe Fields, B. D., Freese, K., \& Graff, D. S. 1998, NewA, 3, 347

\refe Fukugita, M., Hogan, C. J., \& Peebles, P. J. E. 1998, ApJ, 503, 518

\refe Gates, E. I., Gyuk, G., \& Turner, M. S. 1995, Phys. Rev. Lett. 74, 3724

\refe Honma, M. \& Kan-ya, Y. 1998, ApJ, 503, L139

\refe Jakobsen, P. 1998, A \& A, 331, 61  

\refe Mo, H. J., \& Miralda-Escud\'{e}, J. 1996, ApJ, 469, 589

\end{document}